\documentclass{iau} 
\usepackage{graphicx}

\title[Brightness temperatures of galactic masers] 
{Brightness temperatures of galactic masers observed in the RadioAstron project}

\author[Nadya Shakhvorostova et al.]   
{Nadezhda N. Shakhvorostova$^1$,
Alexey V. Alakoz$^1$ \\
\and Andrej M. Sobolev$^2$}

\affiliation{$^1$Astro-Space Center of Lebedev Physical Institute, \\
84/32 Profsoyuznaya st., Moscow, GSP-7, 117997, Russia,\\
email: {\tt nadya@asc.rssi.ru, alexey.alakoz@gmail.com} \\[\affilskip]
$^2$Ural Federal University, 19 Mira street, Ekaterinburg, 620002, Russia \\
email: {\tt Andrej.Sobolev@urfu.ru}}

\pubyear{2017}
\volume{336}  
\jname{Astrophysical Masers: Unlocking the Mysteries of the Universe}
\editors{A. Tarchi, M.J. Reid \& P. Castangia, eds.}
\begin{document}

\maketitle

\begin{abstract}
We present estimates of brightness temperature for 5 galactic masers in star-forming regions detected at space baselines. Very compact features with angular sizes of $\sim$23$-$60~$\mu$as were detected in these regions with corresponding linear sizes of $\sim$4$-$10 million km. Brightness temperatures range from 10$^{14}$ up to 10$^{16}$~K.
\end{abstract}

\firstsection 
\section{Maser observations in the RadioAstron project}
Galactic masers have been observed in the RadioAstron (RA) project during 6 year of operation since the launch in July 2011~[\cite{raes}]. The satellite is equipped with receivers allowing observations of strong maser lines at 22235, 1665 and 1667 MHz. The space interferometer provides a record angular resolution up to 7~$\mu$as at 22 GHz. So, we can put tight limits on the sizes of very compact maser spots, estimate their brightness temperatures and, thus, obtain important parameters for maser models.

The sensitivity of the RA together with the 100-m Effelsberg RT at 22 GHz is $\sim$10~Jy (at 6$\sigma$) with a coherent integration time $\sim$600 sec and typical line width $\sim$0.4~km/s. Observations carried out for H$_2$O masers on the RA indicate in the most cases only a small contribution from ultra-compact components to the total flux of the separate spatial-kinematic features. Thus, the W3~IRS~5, observed on a baseline of 2.5~ED, showed a visibility function amplitude of only 1\% of the total flux density [\cite{sasb2017}].

Such super-compact H$_2$O features were successfully detected in 7 galactic star-forming regions regions and in 2 extragalactic masers in NGC4258 and NGC3079. Current statistics of RA observations can be found in~[\cite{sasb2017}]. In present work we consider 5 galactic H$_2$O masers and obtain the upper limit of the angular size of the most compact components and the lower limit of the brightness temperature. We used the data processed on the ASC software correlator for the RA mission [\cite{corr}].


\section{Brightness temperatures from the interferometric visibilities}
\label{tbsize}

Normally, brightness temperature T$_b$ can be obtained from imaging with a long period of observations and many telescopes involved. But there are a lot of short observations ($\sim$1~hour) in the early RA maser survey with a few baseline sets, about 3~to~6. In this case it is possible to estimate brightness temperature of a source using some assumptions. Thus, without a priory information about brightness distribution, we may use a circular Gaussian and estimate T$_b$ and size of a source as proposed by \cite{lobanov}:

\begin{equation}
\label{tb_eq}
T_{\rm b} = \frac{\pi}{2k}\frac{B^2 V_0}{\ln(V_0/V_q)} [K],
\end{equation}

\noindent where V$_q$ is the visibility amplitude, V$_0$ is the space-zero visibility, $q=B/\lambda$~-- single spatial frequency. It was shown in~[\cite{lobanov}]~that T$_b$ is at its lowest when $V_0/V_q = e$. This provides the minimal brightness temperature given the baseline length and correlated flux obtained from data processing using PIMA package [\cite{petrov}]:
 

\begin{equation}
\label{tbmin_eq}
T_{\rm b,min} \approx 3.09\left(B [{\rm km}]\right)^2\left(V_q [{\rm mJy}]\right) [K].
\end{equation}

\section{Results and conclusions}
\label{results}

Results of our calculations of T$_b$ and T$_{\rm b,min}$ according to \ref{tb_eq} and \ref{tbmin_eq} are given in the Table~\ref{results_table}. Columns contain from left to right: (1) source name, (2) RA and DEC coordinates (J2000), (3) baseline in units of Earth diameters (ED), (4) corresponding resolution in $\mu$as (this value can be considered as an upper limit of angular size of the compact feature), (5) T$_{\rm b,min}$ (the lower limit of T$_b$) and (6) T$_b$. 

\underline{Main conclusions.} In star-forming regions very compact maser features with angular sizes of 23$-$60~$\mu$as were observed, which correspond to $\sim$4$-$10 million km. The best linear resolution was obtained for the H$_2$O maser in Orion~-- 4~million~km. The best angular resolution for Galactic masers is 23 $\mu$as for W49~N (the distance is $\sim$11~kpc). Brightness temperatures for the most compact maser features range from 10$^{14}$ to a few of 10$^{15}$~К.

\begin{table}
\caption{Brightness temperature for compact H$_2$O masers observed in the RA project.}
\centering
\begin{tabular}{|c|c|c|c|c|c|c|}
\hline
Source    & RA (J2000)  & DEC (J2000)               & Baseline,   & Resolution, & T$_{\rm b,min}$,     & T$_b$, \\
          & hh mm ss.ss & $\circ$\quad$\prime$\quad$\prime\prime$ & ED          & $\mu$as     & K                    & K      \\
\hline
Orion KL  & 05 35 14.13 & $-$05 22 36.48            & 3.3         & 66          & 1.2$\times$10$^{15}$ & 6$\times$10$^{15}$ \\
\hline
Cepheus A & 22 56 17.97 & 62 01 48.75               & 3.4         & 64          & 1.2$\times$10$^{14}$ & 3$\times$10$^{14}$ \\
\hline
W3 OH     & 02 27 04.84 & 61 52 24.61               & 3.8         & 58          & 2.1$\times$10$^{14}$ & 7$\times$10$^{14}$ \\
\hline
W3 IRS5   & 02 25 40.71 & 62 05 52.52               & 5.4         & 40          & 1.5$\times$10$^{15}$ & 8$\times$10$^{15}$ \\
\hline
W49 N     & 19 10 13.41 & 09 06 12.80               & 9.6         & 23          & 4.5$\times$10$^{14}$ & 3$\times$10$^{15}$ \\ 
\hline
\end{tabular}
\label{results_table}
\end{table}

\acknowledgements

The RadioAstron project is led by the Astro Space Center of the Lebedev Physical Institute of the RAS and the Lavochkin Association of the Russian Federal Space Agency, and is a collaboration with partner institutions in Russia and other countries: IAA RAS, MPIfR, INAF, NRAO, MASTER Robotic Net, ISON and others.

\end{document}